%% file: 21sips_wifi_sensing.tex
\documentclass[conference,10pt]{IEEEtran}

\ifCLASSINFOpdf
\else
\fi
\usepackage{tikz}
\usepackage{pgfplots}
\usepgfplotslibrary{units}
\usetikzlibrary{positioning}
\usetikzlibrary{arrows}

\usepackage{subcaption}
\usepackage{framed}
\usepackage{balance}

\usepackage{amsthm}

\usepackage[inline]{enumitem}

\pgfplotsset{grid style={gray!50}}
\pgfplotsset{minor grid style={gray!50}}

\usepackage{array}
\usepackage{booktabs}
\usepackage{color}

\definecolor{Set1-7-1}{RGB}{228,26,28}
\definecolor{Set1-7-2}{RGB}{55,126,184}
\definecolor{Set1-7-3}{RGB}{77,175,74}
\definecolor{Set1-7-4}{RGB}{152,78,163}
\definecolor{Set1-7-5}{RGB}{255,127,0}
\definecolor{Set1-7-6}{RGB}{166,86,40}
\definecolor{Set1-7-7}{RGB}{0,0,0}
\definecolor{babyblue}{rgb}{0.54, 0.81, 0.94}

\usepackage[cmex10]{amsmath}
\usepackage{amssymb,amsthm}
\usepackage{mathrsfs}
\usepackage{amsbsy}
\usepackage{steinmetz}
\usepackage{nicefrac}
\usepackage{soul}
\usepackage{multirow}
\usepackage{graphicx}
\usepackage{epstopdf}
\usepackage{amsfonts}
\usepackage{xcolor}

\usepackage[numbers]{natbib}

\makeatletter
\def\testclr#1#{\@testclr{#1}}
\def\@testclr#1#2{{\fboxsep\z@\fbox{\colorbox#1{#2}{\phantom{XX}}}}}

\makeatother

\usepackage{glossaries}
\newacronym{cdf}{CDF}{cumulative distribution function}
\newacronym{mse}{MSE}{mean-squared error}
\newacronym{bs}{BS}{base station}
\newacronym{rssi}{RSSI}{received signal strength indicator}
\newacronym{nlos}{NLOS}{non-line-of-sight}
\newacronym{sfo}{SFO}{sampling frequency offset}
\newacronym{cfo}{CFO}{carrier frequency offset}
\newacronym{pbd}{PBD}{packet boundary detection}
\newacronym{csi}{CSI}{channel state information}
\newacronym{cir}{CIR}{channel impulse response}
\newacronym{ofdm}{OFDM}{orthogonal frequency-division multiplexing}
\newacronym{awgn}{AWGN}{additive white Gaussian noise}
\newacronym{agc}{AGC}{automatic gain control}
\newacronym{boi}{BoI}{band-of-interest}
\newacronym{mad}{MAD}{mean absolute deviation}
\newacronym{psd}{PSD}{power spectral density}
\newacronym{dw}{DW}{discrete wavelet}
\newacronym{dft}{DFT}{discrete Fourier transform}
\newacronym{idft}{IDFT}{inverse discrete Fourier transform}

\title{ComplexBeat: Breathing Rate Estimation from Complex CSI 
}

\author{\IEEEauthorblockN{
Sitian Li\IEEEauthorrefmark{1},  
Andreas Toftegaard Kristensen\IEEEauthorrefmark{1},
Andreas Burg\IEEEauthorrefmark{1},
Alexios Balatsoukas-Stimming\IEEEauthorrefmark{2}
}
\IEEEauthorblockA{\IEEEauthorrefmark{1}Telecommunication Circuits Laboratory, \'{E}cole polytechnique f\'{e}d\'{e}rale de Lausanne, Switzerland\\
\IEEEauthorrefmark{2}Eindhoven University of Technology, the Netherlands\\
}
}

\begin{document}

%\title{Estimating Respiration rate based on selected component in CSI}
%\title{Extricate CSI Components for Respiration Rate Estimation }
%\title{Extracting CSI Components for Breathing Rate Estimation}

\maketitle
% Adjust the size of the footnote number to make it less prominent
\renewcommand{\thefootnote}{\textcolor{white}{\arabic{footnote}}} % Makes footnote number invisible
% Manually add the footnote below the title without a number
\footnotetext{\textcopyright\ 2021 IEEE. This work has been accepted for publication in the 2021 IEEE Workshop on Signal Processing Systems (SiPS). The final published version is available at DOI: \texttt{10.1109/SiPS52927.2021.00046.}}
\footnotemark[0] % This prevents a number from being generated

\begin{abstract}
In this paper, we explore the use of channel state information (CSI) from a WiFi system to estimate the breathing rate of a person in a room. 
%We explain that several CSI phase components are more related and more sensitive to human chest displacement than other components, depending on CSI representation, the environment, and the position and orientation of the person.
In order to extract WiFi CSI components that are sensitive to breathing, we propose to consider the delay domain channel impulse response (CIR), while most state-of-the-art methods consider its frequency domain representation.
One obstacle while processing the CSI data is that its amplitude and phase are highly distorted by measurement uncertainties. 
We thus also propose an amplitude calibration method and a phase offset calibration method for CSI measured in orthogonal frequency-division multiplexing (OFDM) multiple-input multiple-output (MIMO) systems. 
%We provide pipeline systems to process CSI data and integrate our calibration and selection scheme. 
Finally, we implement a complete breathing rate estimation system in order to showcase the effectiveness of our proposed calibration and CSI extraction methods.
%By selecting and filtering the calibrated complex-valued CSI components in the delay domain, our system achieves high accuracy in breathing rate estimation.
\end{abstract}


\IEEEpeerreviewmaketitle

\input{sec/intro_new.tex}
\input{sec/preliminaries_new.tex}
\input{sec/system_new.tex}

\input{sec/experiment_new.tex}

\input{sec/conclusions.tex}
%\input{sec/equations.tex}

\section*{Acknowledgment}
This research has been kindly supported by the Swiss National Science Foundation under Grant-ID 182621.


\balance
\bstctlcite{IEEEexample:BSTcontrol}
\bibliographystyle{IEEEtran}
\bibliography{21sips_wifi_sensing}

\end{document}

%% file: sec/intro_new.tex
\section{Introduction}
% introducation
The observation of vital signs is an essential part of health monitoring, both for hospitalized patients and healthy people in daily life.
Breathing rate, for instance, is an early and extremely good indicator of physiological conditions such as hypoxia (low levels of oxygen in the cells) and hypercapnia (high levels of carbon dioxide in the bloodstream)~\cite{rolfe_importance_2019}.
%Several types of devices have been developed to measure vital signs such as breathing rate and heart rate. 
A polysomnograph (PSG) is a device designed for sleep studies, which consists of belts around the chest and upper abdominal wall for monitoring the body movements from breathing~\cite{pagel_polysomnography_2014}.
Compared to such intrusive or attached devices, non-contact radio frequency (RF) sensors can observe vital signs using the fact that RF signal transmission is influenced by small changes in the environment, without introducing discomfort to users.\par
Initial systems for wireless health monitoring were based on radar, which sends pulses that are reflected and estimates the round-trip time which corresponds to the distance between the sensor and the human body for monitoring the human chest movement during breathing~\cite{costanzo_software_defined_2019, droitcour_signal_noise_2009}. 
To deal with multipath channels,  frequency modulated carrier wave (FMCW) radars were employed by~\cite{anitori_fmcw_2009, peng_portable_2017}. \par
%However, the multipath environment introduces additional reflections which cannot be distinguished from the wave reflected from the human chest.
%To deal with this issue in the channel,  frequency modulated carrier wave (FMCW) radars were employed by \cite{anitori_fmcw_2009, peng_portable_2017} to build an FMCW system for separating received waves from various reflectors located at different ranges. 
%By extracting the FMCW received from a specific distance range, the chest displacement is measured and the breathing rate is further calculated \cite{adib_smart_2015}. 
%The work of \cite{ahmad_vital_2018} shows the feasibility of measuring vital signs of multiple people by multiple FMCW antenna pairs. \par
%Over multiple receive channels, objects are separated in the range-azimuth angle plane and the respiration rates are measured separately.\par
%Radar systems can measure the object displacement in the channel with high accuracy thanks to their large bandwidth. 
%For example, FMCW radar covers a bandwidth of $1.79$ GHz \cite{adib_smart_2015}. 
%However, the high cost of FMCW hardware makes these solutions impractical for everyday home-usage.
%In comparison, devices that occupy less bandwidth are more common in households. 
%Narrow-band devices that measure \Gls{rssi} are employed in \cite{abdelnasser_ubibreathe_2015} for breathing rate monitoring. 
%The estimation accuracy of \gls{rssi}-based methods, however, highly depends on the orientation and position of the person. \par 
Relying on WiFi is another, often more convenient, solution for wireless sensing of vital signs since WiFi transceivers are already widely deployed. 
Commercial WiFi employs \gls{ofdm} for signal transmission, which decomposes the channel into multiple subcarriers. 
By sending pilots, the \gls{cir} can be measured, which provides magnitude and phase information of the different signal propagation paths. 
%An intuitive method is to track the delay of the path reflected by the human chest. 
%However, the delay resolution of WiFi with $40$ MHz bandwidth is very poor, i.e., $25$~ns, corresponding to a low distance resolution of $7.5$~m.
%Consider two paths in the channel with length $3$m and $6$m, the delay difference is only $10$ns. 
%It is very difficult to separate the two paths with a resolution of $25$ns.
%Consider two paths in the channel with a length of several meters, the delay difference is only a few nanoseconds. 
%It is very difficult to separate the two paths with such a small resolution.
%Fortunately, it is still possible to measure the changing rate of the chest displacement by the channel state information (CSI) of WiFi \gls{ofdm}.
Existing vital signs sensing methods consider the frequency domain representation of the \gls{cir} to estimate the chest displacement as in~\cite{Liu_Tracking_2015}.
This representation, composed of multiple subcarriers, is readily available in \gls{ofdm} systems and is typically referred to as \gls{csi} in the corresponding computer science literature.  

%As the \gls{cir}  represents the delay domain of the channel, the discrete Fourier transform (DFT) of \gls{cir} is the spectral domain of the channel, which is CSI.
%The change of one path delay in \gls{cir} influence both amplitude and phase components in the CSI. 
%Thus, we can observe the displacement of the human chest that causes path delay change by observing CSI from \gls{ofdm} WiFi. \par 
The survey in~\cite{ma_wifi_2019} lists a wide range of works on sensing with WiFi.  
Recent works on breathing rate estimation based on \gls{ofdm} WiFi can be broadly grouped into two categories.
%\textbf{CSI amplitude-based respiration rate estimation:} 
In the first category, the breathing rate is estimated by observing periodic \gls{csi} amplitude changes in subcarriers~\cite{Liu_Tracking_2015, lee_design_2018, hillyard_comparing_2018, liu_wi_sleep_2014}. 
%However, the CSI amplitude suffers from uncertainty caused by power control, gain control and noise \cite{zhuo_perceiving_2017}.
%and the impact of small distance variations on power are extremely small 
%\textbf{CSI phase-based respiration rate estimation:}
%The other category is CSI phase-based respiration rate estimation.
In the second category, the phase of the \gls{csi} is used for breathing rate estimation. 
The work in~\cite{wang_phasebeat_2017}, nick-named PhaseBeat,  introduced a method to estimate breathing rate information from the phase difference between two antennas of a multiple-input multiple-output (MIMO) WiFi device with multiple antenna pairs.
Moreover, FullBreathe~\cite{zeng_fullbreathe_2018} and ResBeat~\cite{wang_resilient_2020} estimate breathing rate based on both \gls{csi} amplitude and phase, where one component is complementary to the other, depending on the position of the person. 
However, these methods treat the \gls{csi} amplitude and phase separately as two independent, real-valued measurements, where the complex-valued nature of the CSI is ignored.
Yet, considering the amplitude and phase as a single complex-valued measurement is more natural as treating them independently:
On the one hand, the \gls{idft} of the complex-valued \gls{csi} is the \gls{cir}, in which only the bin with a certain delay, depending on the position of the person, is relevant to the breathing movement.
This component can then be isolated.
On the other hand, since both amplitude and phase of the \gls{csi} contain the information of breathing rate, it is easier to directly observe the periodicity in the complex-valued \gls{csi} stream. \par
%Unfortunately, prior works \cite{wang_phasebeat_2017, zeng_fullbreathe_2018} also observe that the sensitivity of the phase to the breathing movement (depending on the position and orientation of the person) and the impact of the \gls{awgn} on the phase varies strongly between subcarriers. 
%There is fading for subcarrier frequencies in the multipath channel, which means that for some carrier frequencies, the CSI amplitudes are very small. 
%As we shall show later in Section \ref{sec:pre_select}, the amplitude and phase of those subcarriers are very sensitive to \gls{awgn}, which makes the estimation unreliable, compared to other subcarriers with high amplitude.
%Furthermore, since subcarriers correspond to different wavelengths, depending on the position and orientation of the person, subcarrier phase components show different sensitivity to the chest displacement. 
%Therefore, it is essential to merge multiple subcarriers or select the best subcarrier component, which is not only sensitive to the human chest movement, but also robust against \gls{awgn}.
%Existing selection methods employ variance-based \cite{ wang_human_2016} or absolute deviation-based \cite{ wang_phasebeat_2017} criteria.
%Those methods select the subcarrier that shows the highest deviation over time. 
%Some works select the subcarrier by the breathing-to-noise ratio (BNR). They first find the FFT bin with maximal energy within the human breathing range and then compute the ratio between the enegy on that bin and the energy sum of all FFT bins \cite{ zeng_fullbreathe_2018, yue_extracting_2018}.\par
Unfortunately, each \gls{csi} snapshot is affected by a different amplitude distortion caused by the \gls{agc}~\cite{Chen_TRBreath_2017} and a different phase distortion caused by offsets such as \gls{sfo}, \gls{cfo}, and \gls{pbd} error~\cite{ma_wifi_2019}. 
%Such distortions cannot be neglected. 
%Consider the case where a WiFi device is operating in the $5$ GHz band: a small PBD error of $0.1$~ns leads to a phase offset of $\pi$ that crosses half of the feasible region $[0, 2\pi]$ of the phase.
Therefore, a good amplitude and phase calibration for removing measurement artifacts is necessary for estimating the breathing rate from the \gls{csi}.
The works in~\cite{Liu_Tracking_2015, hillyard_comparing_2018} employ low-pass filters on the \gls{csi} amplitude along time to reduce the influence of \gls{agc}.
In~\cite{gao_crisloc_2021}, the \gls{agc} is eliminated by scaling the \gls{csi} amplitude by the sum of CSI amplitude squared over all subcarriers.
To cancel the influence of phase offsets, the works of~\cite{wang_phasebeat_2017, wang_resilient_2020} employ a receiver with multiple antennas and use the phase of one of the antennas as a reference to be subtracted from other antennas. 
Another method is to use the linear component in the phase with respect to subcarrier indices as a reference to eliminate phase distortion~\cite{Wang_PhaseFi_2015}. \par
%However, by subtracting the phase component from one antenna, the complex property of the CSI is withdrawn. Transforming the CSI into \gls{cir} and further data processing in delay domain is not possible.

\subsection*{Contributions}
In this paper, we introduce a novel \gls{csi} calibration method for both amplitude and phase distortions. 
We calibrate the amplitude of the \gls{csi} based on the amplitude of the dominant path in the \gls{cir}.
The phase is calibrated by the linear components of the  \gls{csi} phase of the adjacent receiver antenna.
Both the complex-valued \gls{csi} and \gls{cir} are then used to estimate the breathing rate, after carefully selecting the most relevant signal components.
%The calibrated \gls{csi} is further processed in the delay domain to extract the relevant component to breathing movement.
We implement two different breathing rate estimation systems with tunable parameters and test them on the same publicly available data set~\cite{hillyard_experience_2018}, to evaluate our proposed calibration and extraction methods.

%\subsection*{Outline}
%In section \ref{sec:Preliminaries}, we present the preliminaries for estimating breathing rate from CSI phase, criteria for selecting CSI phase components that are sensitive to breathing as well as robust to noise and a calibration method for CSI to eliminate phase offset.
%We introduce three system pipeline structures in section \ref{sec:System}, and we compare the performance between systems based on the data set in section \ref{sec:experiment}.

%% file: sec/preliminaries_new.tex
\section{Preliminaries}
In this section, we describe the structure of the \gls{csi} and the \gls{csi} phase distortion.
Moreover, we explain the phase difference method described in~\cite{wang_phasebeat_2017} as basis for the proposed improvements described in Section~\ref{sec:complexbeat}.
Finally, we discuss the existing subcarrier selection and periodicity detection methods for breathing rate estimation using \gls{csi} data.
\label{sec:Preliminaries}
\subsection{CSI in WiFi MIMO OFDM Systems}
Consider an OFDM system operating over a multipath channel.
The transmitter sends training information and the receiver extracts the complex baseband \gls{csi} that can be described as
%The source signal in the time domain is denoted as $s(t)$. 
%After up-conversion to the carrier frequency, the transmitted signal is written as $\tilde{s}(t) = s(t)e^{j2\pi f_c t}$, where $f_c$ is the center carrier frequency.
%At the receiver side, the reflected signal along path $p$ arrives at the receiver with delay $\tau_p$, and accumulates with other paths:
%\begin{align}
%\tilde{r}(t) = \sum_{p=1}^{P} A_p s(t-\tau_p) e^{j2\pi %f_c(t-\tau_p)}
%\end{align}
%where $P$ is the number of paths, $A_p$ is the attenuation on path $p$.
%After down-conversion at the receiver side, here we assume the transmitter and receiver are perfectly synchronized and there is no phase noise, the received signal is:
%\begin{align}
%r(t) 
%&= \tilde{r}(t) e^{j2\pi f_c(-t)} \notag \\
%&= \sum_{p=1}^{P} A_p s(t-\tau_p) e^{j2\pi f_c(-\tau_p)}
%\end{align}
%By comparing source signal $s(t)$ and received $r(t)$, the \gls{csi} can be written as: 
%\begin{align}
%h(t) = \sum_{p=1}^{P} A_p \delta(t-\tau_p) e^{j2\pi f_c(-\tau_p)}
%\end{align}
%where $\delta$ is Dirac delta function.
%If we calculate the DFT of $h(t)$, we transform from delay domain to spectral domain of the channel, which is: 
%\begin{align}
%H(f) = \sum_{p=1}^{P} A_p e^{-j 2 \pi (f-f_c) \tau_p}
%\end{align}
%And by replacing $f$ with $\Delta f \tilde{m}$, we obtain the discrete version of $H(f)$, which is \gls{csi}:
\begin{align}
H_{m} = \sum_{p=1}^{P} A_p e^{-j 2 \pi (\Delta f m+f_c) \tau_p},
\label{equ:pre_csi_multi_path}
\end{align}

%where $\Delta f$ is the subcarrier spacing, $m$ is subcarrier index, 
where $P$ is the number of paths, $A_p$ and $\tau_p$ are the attenuation and delay of path $p$, respectively, $m$ is the subcarrier index, $\Delta f$ is the subcarrier spacing and $f_c$ is the carrier frequency.
Note that $\Delta f m$ is the baseband carrier frequency, which is between $-\frac{B}{2}$ and $\frac{B}{2}$, where $B$ is the bandwidth.
In an OFDM communication system, the \gls{csi} has $M$ entries, and each entry is a function of $\tau_p$, which can be used to estimate the change of $\tau_p$ over time.
%In an OFDM communication system, the \gls{csi} is commonly represented by a vector $\mathbf{H} = [H_1, H_2, \cdots, H_M]^T$, where $M$ is the number of subcarriers.
%Each entry of $\mathbf{H}$ is a complex number consist of an amplitude $|H_m|$ and a phase $\angle H_m$, representing the channel condition for each subcarrier $m$.
%for the $m$ th subcarrier is described as a complex number $H_m$ consists of the magnitude $|H_m|$ and angle $\angle H_m$ components.
%As shown in (\ref{equ:pre_csi_multi_path}), $H_{m}$ is the superposition of complex numbers $A_p e^{-j 2 \pi (\Delta f m+f_c) \tau_p}$ on each path $p$ . 

%We use $m$ to denote the subcarrier indices, instead of elements in such that $\Delta f \tilde{m}$ is the baseband carrier frequency.

%The vector formed by all $M$ subcarries' $H_m$ is the \gls{csi} $\mathbf{H} = [H_1, H_2, \cdots, H_M]^\top$.
For the WiFi MIMO OFDM case, where there is more than one transmitter (TX) / receiver (RX) antenna pair, the \gls{csi} can be written as a tensor $H_{i, j, m} \in \mathbb{C}^{N_T\times N_R\times M}$, where $N_T$ and $N_R$ are the number of TX and RX antennas, respectively. 
To observe the change in the channel, multiple \gls{csi} snapshots are obtained. 
A \gls{csi} sample set is written as a four-dimensional tensor $H_{i, j, m}[t] \in \mathbb{C}^{T\times N_T\times N_R\times M}$, where $T$ is the number of snapshots.

\subsection{CSI Phase Distortion and RX Phase Difference}
\label{sec:CSI_phasediff}
When transmitter and receiver are not synchronized and phase noise is present, \gls{pbd} error, \gls{sfo}, and \gls{cfo} have an influence on the phase of the \gls{csi}.
The phase of the \gls{csi} sample set, including these distortions can be described as~\cite{speth_optimum_1999, zhuo_identifying_2016, xie_precise_2019}
\begin{align}
\angle \tilde{H}_{i, j, m}[t] = \angle H_{i, j, m}[t] + (\xi^{d}_i[t]+\xi^{s}_i[t]) m + \xi^{c}_i[t] + \gamma_i[t],
\label{equ:pre_phase_h1}
\end{align}
where $\angle \tilde{H}_{i, j, m}[t]$ is the distorted \gls{csi} phase value for TX antenna~$i$, RX antenna~$j$ and subcarrier index~$m$ at snapshot~$t$.
The variable $\gamma_i$ is the initial phase offset due to the phase-locked loop (PLL), $\xi^{d}_i$, $\xi^{s}_i$, and $\xi^{c}_i$ are the phase errors from the \gls{pbd} error, \gls{sfo}, and \gls{cfo}, respectively for TX antenna~$i$~\cite{speth_optimum_1999, zhuo_identifying_2016, xie_precise_2019}. 
The offset values are identical for all RX antennas on the same receiver at the same snapshot $t$.
%The phase components of CSI from TX antenna $i$ to RX antenna $j$ is
%\begin{align}
%\angle \tilde{H}_{i, j,\tilde{m}} = \angle H_{i, j,\tilde{m}} + ((\xi^{d}_i+\xi^{s}_i) \tilde{m}_i + \xi^{c}_i + \gamma_i )
%\label{equ:pre_phase_h1}
%\end{align}
In order to eliminate the unknown phase component $(\xi^{d}_i[t]+\xi^{s}_i[t]) m + \xi^{c}_i[t] + \gamma_i[t]$, the work of~\cite{wang_phasebeat_2017} proposed to subtract $\angle \tilde{H}_{i, j',m}[t]$ from $\angle \tilde{H}_{i, j,m}[t]$, where $j'\neq j$, and obtain the RX phase difference $\angle \tilde{H}_{i, k,m}[t]$
\begin{align}
\angle \tilde{H}_{i, k,m}[t] =\angle H_{i, j,m}[t] - \angle H_{i, j',m}[t].
\label{equ:pre_phase_diff}
\end{align}
%By taking multiple snapshots of $\angle \tilde{H}_{i, k,\tilde{m}}$ over time, we obtain $\angle \tilde{H}_{i, k,\tilde{m}}[t]$. 
According to~\cite[Lemma 1]{wang_phasebeat_2017}, when the wireless signal is reflected from the chest of a person with a breathing frequency $f_r$, the true phase of the reflected signal at any antenna of the receiver is also periodic with the same frequency $f_r$.
The phase difference $\angle \tilde{H}_{i, k,m}[t]$, as the linear combination of $\angle H_{i, j, m}[t]$ and $\angle H_{i, j', m}[t]$, contains this periodic component of $f_r$ as well.
However, the drawback of this calibration is that the resulting \gls{csi} with~(\ref{equ:pre_phase_diff}) as its phase component no longer reflects the physical \gls{cir}.  

\subsection{Signal Component Selection}
\label{sec:sub_sel}
Even though it has been shown that the phase of the \gls{csi} is periodic with $f_r$, several works~\cite{wang_phasebeat_2017, zeng_fullbreathe_2018, wang_resilient_2020} have shown that the phase component of different subcarriers shows different sensitivity to breathing movement.
To select a suitable subcarrier, a \gls{mad}-based method is proposed in~\cite{wang_phasebeat_2017} in which three subcarriers with the largest maximum \gls{mad} on measured phase difference are pre-selected. 
Then the subcarrier with the median \gls{mad} value among the pre-selected three candidates is finally selected.

Another possible selection method is based on \gls{boi} analysis. 
%For this analysis, the \gls{csi} phase or phase difference, for antenna indices $i$, $j$ and subcarrier index $m$, are denoted as $\Phi_{i, j, m}[t]$, as the \gls{csi} feature candidates.
%In the following, we denote the phase on antenna $i$, $j$ and subcarrier $m$ or phase difference value with index $i, k$ and $\tilde{m}$ over snapshot $t$ as $\Phi_{i, j, \tilde{m}}[t]$.
The corresponding selection criterion is based on the \gls{psd} $P_{i, j,m}[f]$ along the snapshot axis of any measured and calibrated quantity, which can be either the phase difference of the \gls{csi} or the complex-valued \gls{csi} or \gls{cir} directly.
From the \gls{psd}, a score $S$ is obtained as 
\begin{align}
S_{i, j,m} = \frac{\sum_{f_r^l \leq |f| \leq f_r^h } |P_{i, j,m}[f]|^2}{\sum_{|f| > f_r^h} |P_{i, j,m}[f]|^2},
\end{align}
where $f_r^l$ and $f_r^h$ are the lower and upper limits of the human breathing rate $f_r$.
The band between $f_r^l$ and $f_r^h$ is the \gls{boi}.
The larger $S_{i, j, m}$ is, the more likely it is that the snapshot sequence with $i$, $j$ and $m$ is sensitive to breathing movement.
%This selection method can be applied to complex-valued measurements as well, i.e., we compute a score for the corresponding \gls{csi} $H_{i, j,m}[t]$ or the \gls{cir} $h_{i, j,\tau}[t]$ instead of $\Phi_{i, j, m}[t]$.

\subsection{Breathing Rate Estimation}
\label{sec:rate_est}
After selecting the best candidate, denoted with indices $i^*$, $j^*$ and $m^*$, we detect the breathing rate from the signal component (e.g., $\hat{H}_{i^*, j^*, m^*}[t]$) along snapshots $t$.
In the work of~\cite{Liu_Tracking_2015} and~\cite{wang_phasebeat_2017}, a peak detection method is used to extract breathing rate information. 
The local peaks along $t$ are first detected, and fake peaks are removed using~\cite[Algorithm 1]{Liu_Tracking_2015}.
After peak detection,  all peak-to-peak intervals are averaged to obtain the periodicity of the breathing signal~\cite{wang_phasebeat_2017}.

However, our analysis shows that this peak detection is not robust against noise in the signal change along time. Once there is one peak suppressed by noise or interference, there will be a large error in the breathing rate estimation.
Thus, we also consider the \gls{psd}-detect method described in~\cite{hillyard_comparing_2018}.
Specifically, we compute the \gls{psd} of the selected signal component with higher resolution by padding zeros to the end of the measurement (along the time axis). 
The frequency at which the \gls{psd} is maximum is returned as the estimated breathing rate.
This \gls{psd}-detect method can also be applied to the complex-valued waveform over~$t$.

\section{ComplexBeat}
\label{sec:complexbeat}
In this section, we first introduce a new \gls{csi} calibration method that preserves most of the ground truth phase information in the original, uncalibrated \gls{csi}.
Then we further process the calibrated complex \gls{csi} in the delay domain.
%We propose a novel calibration method that is used to eliminate amplitude and phase distortion from the measurement caused by \gls{agc} and receiver phase offsets.
%The ground truth \gls{csi} is 
%\begin{align}
%H_{i, j, m}[t] = | H_{i, j, m}[t] | e^{j \angle H_{i, j, m}[t]}, 
%\end{align}
%and the corresponding \gls{cir} is 
%\begin{align}
%h_{i, j, \tau}[t] = | h_{i, j, \tau}[t] | e^{j \angle h_{i, j, \tau}[t]}, 
%\end{align}
The uncalibrated \gls{csi} at snapshot $t$ is the ground truth \gls{csi} multiplied by an \gls{agc} term $A_j^{\text{AGC}}$~\cite{Chen_TRBreath_2017, zhuo_perceiving_2017} and a phase offset term~\cite{speth_optimum_1999, zhuo_identifying_2016, xie_precise_2019}
\begin{align}
\tilde{H}_{i, j, m}[t] = A_j^{\text{AGC}}[t] H_{i, j, m}[t] e^{j2\pi ((\xi^{d}_i[t]+\xi^{s}_i[t]) m + \xi^{c}_i[t] + \gamma_i[t])}.
\end{align}

\subsection{CSI Amplitude Calibration: Dominant Path Approach}
\label{sec:pre_calib_amp}
In order to cancel the amplitude distortion caused by the \gls{agc}, we introduce an amplitude calibration based on the dominant component in the \gls{cir}. 
Since the $A_j^{\text{AGC}}[t]$ is the same for each subcarrier $m$, the distorted amplitude of the measured \gls{cir} is $|\tilde{h}_{i, j, \tau}[t]| = A_j^{\text{AGC}}[t] |h_{i, j, \tau}[t]|$. 
Now we assume that \begin{enumerate*}[label=(\roman*)]
\item the delay spread is small, i.e., that there is a dominant path between the transmitter and the receiver, 
\item the path reflected on the human chest is the only dynamic path, and
\item that the breathing movement has very little impact on the amplitude of the dominant path in the \gls{cir}.
\end{enumerate*}
We first determine the dominant path by finding $\tau_{i, j, t}^* = \arg\max_{\tau} |\tilde{h}_{i, j, \tau}[t]|$. 
Then we calculate the calibrated \gls{csi} as
\begin{align}
|\hat{H}_{i, j, m}[t]| 
& = \left(\frac{1}{2D+1}\sqrt[]{\sum_{\scriptscriptstyle |\tau-\tau_{i, j, t}^*|\leq D} |\tilde{h}_{i, j, \tau}[t]|^2} \right)^{-1} |\tilde{H}_{i, j, m}[t]| \notag \\
& = \left(\frac{1}{2D+1}\sqrt[]{\sum_{\scriptscriptstyle |\tau-\tau_{i, j, t}^*|\leq D} |h_{i, j, \tau}[t]|^2} \right)^{-1} |H_{i, j, m}[t]|,
\label{equ:pre_calib_amp}
\end{align}
with a small $D$ that is selected such that the term, $\sum_{\scriptscriptstyle |\tau-\tau_{i, j, t}^*|\leq D} |h_{i, j, \tau}[t]|^2$ covers the power in the dominant path.
With the assumption that the breathing movement has very little impact on the amplitude of the dominant path and that the other paths in the environment are stable, the calibrated \gls{csi} amplitude $|\hat{H}_{i, j, m}[t]|$  is the ground truth \gls{csi} amplitude $|H_{i, j, m}[t]|$ multiplied by an almost constant term that is not changing over time $t$.

\subsection{CSI Phase Calibration: Linear Component Approach}
\label{sec:pre_calib_phase}
Instead of calculating the difference between the phase components on two RX antennas as shown in~\cite{wang_phasebeat_2017}, we introduce a calibration method that preserves most of the phase information of the \gls{csi}. 
%Different from the self-linear calibration method that has been proposed in \cite{7417517}, we calibrate the phase of one RX antenna using the \gls{csi} phase from another RX antenna.
To this end, we exploit that the phase offsets, as shown in~(\ref{equ:pre_phase_h1}), for all RX antennas are identical and are linear functions of the subcarrier index $m$. 
We thus write the term $(\xi^{d}_i[t]+\xi^{s}_i[t]) m + \xi^{c}_i[t] + \gamma_i[t]$ as $\alpha^{\text{off}}_i[t] m + \beta^{\text{off}}_i[t]$, which yields 
\begin{align}
\angle \tilde{H}_{i, j, m}[t] = \angle H_{i, j, m}[t] + (\alpha^{\text{off}}_i[t] m + \beta^{\text{off}}_i[t]).
\end{align}
%Now we assume that the delay spread is small, i.e., that there exists a dominant path between the transmitter and the receiver that has delay $\tau_0$ and amplitude $A_0 \gg A_p$ ($p \neq 0$), and the path reflected on the human chest is the only dynamic path. 
As in Section~\ref{sec:pre_calib_amp}, we assume that the delay spread is small, and that the path reflected on the human chest is the only dynamic path.
%This assumption is not completely realistic, but it provides valuable insight. 
Then, the \gls{csi} phase $\angle H_{i, j, m'}[t]$ of a small subset of adjacent subcarriers $m'$ can also be approximated as linear with respect to $m'$ and the corresponding linear coefficients are denoted as $\alpha'_{i,j}[t]$ and $\beta'_{i,j}[t]$. 
With the influence of phase distortion, the linear trend coefficients along $m'$ of $\angle \tilde{H}_{i, j, m'}[t]$ become $\tilde{\alpha}'_{i,j}[t] = \alpha'_{i,j}[t] + \alpha^{\text{off}}_i[t]$ and $\tilde{\beta}'_{i,j}[t] = \beta'_{i,j}[t] + \beta^{\text{off}}_i[t]$, where $\tilde{\alpha}'_{i,j}[t]$ and $\tilde{\beta}'_{i,j}[t]$ can be obtained from the \gls{csi} phase measurement $\angle \tilde{H}_{i, j, m}[t]$.
%Then we define the term $\text{lin}(\angle \tilde{H}_{j, \tilde{m}'}) := \tilde{\alpha}_j' \tilde{m} + \tilde{\beta}_j'$.
%With the influence of phase distortion, $\text{lin}(\angle \tilde{H}_{j, \tilde{m}'}) = (\alpha_j' + \alpha) \tilde{m} + (\beta_j' + \beta)$.
We then calibrate the \gls{csi} phase on the RX antenna with index~$j+1$ by subtracting $ \tilde{\alpha}'_{i,j}[t] m + \tilde{\beta}'_{i,j}[t]$ from it.
Note that the RX antenna index computation must be performed modulo~$N_R$.
After calibration, the \gls{csi} is written as
\begin{align}
\angle \hat{H}_{i, j+1, m}[t]
=  \angle H_{i, j+1, m}[t] - (\alpha'_{i,j}[t] m + \beta'_{i,j}[t]).
\label{equ:calib01}
\end{align}
%Similarly, we can calibrate $\tilde{H}_{j_a, \tilde{m}}$ with $\text{lin}(\angle \tilde{H}_{j_b, \tilde{m}})$. 
If the signal reflected on the human chest is not on the dominant path for both antennas and the dominant path is static over time, the calibrated $\angle \hat{H}_{i, j+1, m}$ is the true \gls{csi} phase for antenna $j+1$ plus a term, $-j (\alpha'_{i,j}[t] m + \beta'_{i,j}[t])$ that is almost constant across snapshots $t$ for all $m$, as shown in~(\ref{equ:calib01}).
However, if the signal reflected on the human chest is on the dominant path, i.e., the breathing movement affects $\alpha'_{i,j}[t]$ and $\beta'_{i,j}[t]$, the phase of (\ref{equ:calib01}) still contains the term $(\alpha'_{i, j+1}[t] - \alpha'_{i,j}[t]) m  + (\beta'_{i,j+1}[t] - \beta'_{i,j}[t]) $, which still has a periodic component of $f_r$, according to~\cite[Lemma 1]{wang_phasebeat_2017}.
In this case, our method has an advantage compared to the similar self-linear calibration method discussed in~\cite{Wang_PhaseFi_2015}, where the antenna is calibrated by its own linear phase component which causes the phase term that contains useful breathing rate information to vanish.

%From \cite[Lemma 1]{wang_phasebeat_2017}, the phase component of $H_{i, j, \tilde{m}}[t]$ is periodic with the breathing rate $f_r$, thus the $\hat{H}_{i, j, \tilde{m}}[t]$ is also periodic with $f_r$, if the dominant path is static.

\subsection{Delay Domain Detection and Filter}
\label{sec:delay_filter}
Since the amplitude and the phase distortion of the \gls{csi} are eliminated by the calibration while preserving most of the phase information, we can transform the \gls{csi} $\hat{H}_{i, j, m}[t]$ into the delay domain as \gls{cir} $\hat{h}_{i, j, \tau}[t]$.
From the CIR, we find the delay component $\tau^*$, that contains the maximum signal power in the \gls{boi}.

However, the delay resolution of the \gls{cir} is very low due to the narrow bandwidth in \gls{ofdm} WiFi.
The breathing component may therefore be spread across multiple bins in the \gls{cir}.
In order to combine the information in multiple selected bins in the \gls{cir}, we preserve all potentially relevant bins and transform the \gls{cir} back into \gls{csi} representation.
We then select the one most relevant (e.g., according to the BoI criterion) \gls{csi} component instead of a single \gls{cir} component, since all entries of the \gls{csi} are a linear combination of the \gls{cir} bins.
We refer to the selection of a range of  bins in the \gls{cir} as delay filter~(DF).
%The majority of existing \gls{csi}-based breathing estimation methods rely on filtering in the snapshot domain. 
A straight forward approach to determine a range of taps to preserve is based on an upper bound $\tau_{\text{max}}$ for the delay of relevant single-reflection paths on the human chest. 
All \gls{cir} components with a delay $|\tau| >\tau_{\text{max}}$ are set to zero and the resulting filtered \gls{cir} is transformed back into the frequency domain $\hat{H'}_{i, j, m}[t]$,
\begin{align}
\hat{H'}_{i, j, m}[t] = \mathcal{F}_{\tau \rightarrow m}\left(
\begin{cases}
\mathcal{F}^{-1}_{m \rightarrow \tau}(\hat{H}_{i, j, m}[t]), & \text{if}\ |\tau| \leq \tau_{\text{max}} \\
0, & \text{otherwise}
\end{cases}
\right).
\label{equ:delay_filter}
\end{align}
%\begin{figure}[h]
%\input{./pics/tikz_plot_cir_block.tex}
%\end{figure}

%% file: sec/system_new.tex
\section{System Structure}
\label{sec:System}
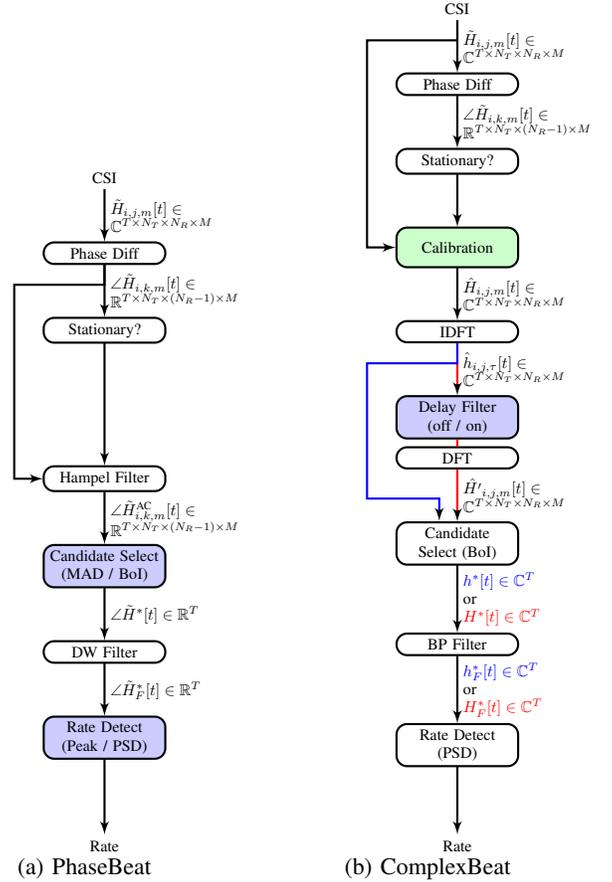
\begin{figure}[t]\centering
\subcaptionbox{
PhaseBeat
\label{fig:sys1}
}
{
\input{./pics/tikz_phase_beat_long_new.tex}
}\quad
\subcaptionbox{
ComplexBeat
\label{fig:sys3}
}
{
\input{./pics/tikz_ours_long_new2.tex}
}
\caption{System Structures}
\label{fig:system}
\end{figure}

In this section, we describe two different breathing rate estimation systems, PhaseBeat~\cite{wang_phasebeat_2017} and our proposed ComplexBeat system, each with different variants. All systems start from a set of \gls{csi} parameter estimates $\tilde{H}_{i, j, m}[t] \in \mathbb{C}^{T\times N_T \times N_R \times M}$.
The corresponding data flows are shown in Fig.~\ref{fig:sys1} for PhaseBeat and in Fig.~\ref{fig:sys3} for ComplexBeat.
Both systems have blocks whose function can be configured.

In the PhaseBeat setup, the candidate selection block and the rate detection block can be configured with different methods. 
By configuring the methods for each block, we obtain the PhaseBeat system as in~\cite{wang_phasebeat_2017} and its two variants:
\begin{enumerate}
\item The PhaseBeat system, which employs the phase difference method and uses the \gls{mad}-based candidate selection method and peak detection method as proposed in~\cite{wang_phasebeat_2017}.
\item PhaseBeat-MAD-PSD, which augments PhaseBeat with \gls{psd}-based rate detection method.
\item PhaseBeat-BoI-PSD, which replaces the \gls{mad}-based selection in PhaseBeat-MAD-PSD with the \gls{boi} selection method.
\end{enumerate}

In ComplexBeat, after implementing our calibration method, we can use either the \gls{cir} or \gls{csi} data representation for further processing.
If the \gls{csi} is used, an optional delay filter is applied.
Different configurations also lead to three ComplexBeat setups for performance comparison:
\begin{enumerate}
\item ComplexBeat-CIR, which implements our calibration method and then applies \gls{cir} for breathing rate estimation, with BoI selection and \gls{psd} rate detection.
\item ComplexBeat-CSI, which modifies ComplexBeat-CIR by using \gls{csi} for breathing rate estimation. The delay filter is turned off.
\item ComplexBeat-CSI-DF, which augments ComplexBeat-CSI by activating the delay filter.
\end{enumerate}

\subsection{PhaseBeat}
\label{sec:phasebeat}
PhaseBeat~\cite{wang_phasebeat_2017} first calculates the \gls{csi} phase difference between two adjacent RX antennas $j$ and $j'\neq j$  as shown in (\ref{equ:pre_phase_diff}). 
%The output is a real-valued phase component array with shape $\mathbb{R}^{T\times N_T\times (N_R-1)\times M}$.

After extracting the \gls{csi} phase difference $\angle \tilde{H}_{i, k, m}$ for $i \in [0, N_T-1]$ and $k \in [0, (N_R - 1)-1]$,  the algorithm determines whether the person is in a stationary state based on a score 
%When the person is lying on the bed without big-ranged movement such as rolling over, this block should make a stationary positive decision, and the breathing rate estimation by the following blocks can be operated. 
\begin{align}
Q = &\frac{1}{N_T \cdot( N_R-1) \cdot M \cdot  T } \times \notag \\
& \sum_{i}  \sum_k \sum_{m}\sum_{t=0}^{T-1} |\angle \tilde{H}_{i, k, m}[t] - \mathbb{E}_t[\angle \tilde{H}_{i, k, m}[t]]|.
\label{equ:stable}
\end{align}
Breathing rate estimation is only performed on sample sets for which $Q<0.18$, since only these are considered as sufficiently stable.

%A phase difference of a sample set, $\angle \tilde{H}_{i, k, m}[t]$ that has been decided as stable is fed in to the candidate selection block. 
%This block decides which candidate is more sensitive to human chest displacement. 
For each stable sample set, we apply a Hampel filter with a window size of $5$~seconds and a threshold of $0.01$ to filter out the DC component as proposed in~\cite{wang_phasebeat_2017}.  
The DC component is subtracted from $\angle \tilde{H}_{i, k, m}[t]$, which leads to a detrended waveform $\angle \tilde{H}^{\text{AC}}_{i, k, m}[t]$.
Then a selection of the most sensitive candidate subcarriers is performed.
PhaseBeat and PhaseBeat-MAD-PSD use the \gls{mad}-based selection method, whereas the PhaseBeat-BoI-PSD applies the \gls{boi}-based method with  $f_r^h = 0.5$~Hz and $f_r^l = 0.2$~Hz.

The selected candidate is then a 1D signal as a function of $t$, denoted as $\angle \tilde{H}^*[t]$.
A \gls{dw} filter is then applied to $\angle \tilde{H}^*[t]$. We calculate the ``db4'' Daubechies decomposition to $\angle \tilde{H}^*[t]$ as in~\cite{wang_phasebeat_2017}, 
\begin{align}
\angle \tilde{H}^*[t]
&= \sum_{k\in \mathbb{Z}} \lambda^{(L)}[k] \phi^{(L)}[t-2^L k] \notag \\
&+ \sum_{l=1}^L \sum_{k\in \mathbb{Z}} \mu^{(l)}[k] \psi^{(l)} [t-2^l k].
\label{equ:db4}
\end{align}
We then only keep the wavelet components whose frequencies are in the range of human breathing rates (from $0$ to $0.5$~Hz), which is the first term in (\ref{equ:db4}). 
The value of $L$ is determined based on the snapshot sampling frequency $f_s$, such that $\frac{f_s}{2^{L+2}} < 0.5 $~Hz $ \leq \frac{f_s}{2^{L+1}}$.
%\begin{align}
%\Phi^*_{\text{LP}}[t] 
%= \sum_{k\in \mathbb{Z}} \lambda^{(L)}[k] \phi^{(L)}[t-2^L k] \notag
%\end{align}
The coefficients $\lambda^{(L)}[k]$ are calculated as $\lambda^{(L)}[k] = \sum_{t\in \mathbb{Z}} \angle \tilde{H}^*[t]  \phi^{(L)}[t-2^Lk] $, where $\phi^{(L)}[t]$ is the Daubechies wavelet on level $L$.

In the end, we calculate the breathing rate using both the peak-detect and the \gls{psd}-detect methods as discussed in Section~\ref{sec:rate_est}.
PhaseBeat uses the peak-detect method. 
PhaseBeat-MAD-PSD and PhaseBeat-BoI-PSD apply the \gls{psd}-detect method with a resolution of $0.001$~Hz. 
The frequency between $0.2$~Hz and $0.5$~Hz for which the \gls{psd} is maximized is chosen as the breathing rate estimate $f_r$. 

\subsection{ComplexBeat}
In this setup, we employ our proposed \gls{csi} calibration as discussed in Section~\ref{sec:pre_calib_amp} and Section~\ref{sec:pre_calib_phase}, for each sample set that is deemed stable according to~(\ref{equ:stable}), using~(\ref{equ:pre_calib_amp}) and~(\ref{equ:calib01}).
Due to frequency selective fading, we select the subcarriers~$m'$ whose amplitude is large to extract $\tilde{\alpha}'_{i, j}[t]$ and $\tilde{\beta}'_{i, j}[t]$,  and calibrate  $\angle \tilde{H}_{i, j+1, m}[t]$ with $-(\tilde{\alpha}'_{i, j}[t] m + \tilde{\beta}'_{i, j}[t])$.
We can continue processing the calibrated data either in its \gls{cir} form (ComplexBeat-CIR, blue thread in Fig.~\ref{fig:sys3}) or in its \gls{csi} form (ComplexBeat-CSI and ComplexBeat-CSI-DF, red thread in Fig.~\ref{fig:sys3}).

ComplexBeat-CIR first obtains the CIR and then uses the \gls{boi} selection method with  $f_r^h = 0.5$~Hz and $f_r^l = 0.2$~Hz, based on the \gls{psd} of $\hat{h}_{i, j, \tau}[t]$ to determine $i^*$, $j^*$, and $\tau^*$ as discussed in Section~\ref{sec:sub_sel}.
We then use the \gls{cir} component $h^*[t]=\hat{h}_{i^*,j^*,\tau^*}[t]$ for further processing.
Both ComplexBeat-CSI and ComplexBeat-CSI-DF operate on the complex-valued \gls{csi}.
In ComplexBeat-CSI-DF, we apply the delay domain filter with $\tau_{\text{max}} = 50$~ns to the complex-valued \gls{csi}, as shown in~(\ref{equ:delay_filter}) in Section~\ref{sec:delay_filter}. 
If the delay filter is off, as configured in ComplexBeat-CSI, the input \gls{csi} is directly copied to the output such that $\hat{H}'_{i, j, m}[t] = \hat{H}_{i, j, m}[t]$.
For both ComplexBeat-CSI and ComplexBeat-CSI-DF, the \gls{csi} candidate with the highest score~$H^*[t] = \hat{H}'_{i^*,j^*,m^*}[t]$ in the \gls{boi} selection is used for further processing.

For all three variants, a band-pass filter is then applied by first transforming the complex series $H^*[t]$ or $h^*[t]$ into their spectrum $P^*[f]$, and setting the values on $|f| < 0.2$~Hz and $|f| > 0.5$~Hz to zero, and then transforming it back to time domain $H^*_F[t]$ or $h^*_F[t]$.
In the end, the \gls{psd}-detection method same as in Section~\ref{sec:phasebeat} is employed for breathing rate detection.
%ComplexBeat-P and ComplexBeat-PD use the phase component of the \gls{csi} $\angle \hat{H}'_{i, j, m}[t]$, and select the candidate by \gls{boi} scheme. 
%The band pass filter is the concatenation of the Hampel filter and the \gls{dw} filter as discussed in Section~\ref{sec:phasebeat}. 
%In the end, the \gls{psd}-based rate detection method is applied to the filtered data.

%ComplexBeat-CD uses the complete \gls{csi} data $\hat{H}'_{i, j, \tilde{m}}[t]$ as a complex input. 

%As discussed in Section~\ref{sec:sub_sel} and Section~\ref{sec:rate_est}, the \gls{boi}-select and \gls{psd}-detect methods can be both applied to complex values since they both process the data in the spectral domain.

%% file: pics/tikz_phase_beat_long_new.tex
\tikzstyle{pipeBox} = [rectangle, draw, fill=white!20, node distance=0.7cm, text width=7em, text centered, rounded corners, minimum height=1em, thick]
\tikzstyle{pipeBoxNull} = [rectangle, draw=none, fill=none, node distance=0.7cm, text width=5em, text centered, rounded corners, minimum height=1em, thick]
\tikzstyle{pipeNexBox1} = [rectangle, draw, fill=green!20, node distance=0.7cm, text width=7em, text centered, rounded corners, minimum height=2.5em, thick]
\tikzstyle{pipeNexBox2} = [rectangle, draw, fill=red!20, node distance=0.7cm, text width=7em, text centered, rounded corners, minimum height=1em, thick]
\tikzstyle{pipeNexBox3} = [rectangle, draw, fill=blue!20, node distance=0.7cm, text width=7em, text centered, rounded corners, minimum height=1em, thick]
\tikzstyle{pipSig} = [rectangle, fill=white, node distance=1.0cm, text width=8em, text centered, rounded corners, minimum height=1em, thick]
\tikzstyle{pipearrow} = [draw, -latex',thick]

\begin{tikzpicture}[thick,scale=0.6, every node/.style={scale=0.6}]

\node [pipSig] (csi) {CSI};
%\node [pipeBox, below=of csi] (buffer)  {Buffer};
\node [pipeBox, below=of csi]  (phase_diff) {Phase Diff};
\node [pipeBox, below=of phase_diff]  (stationary)  {Stationary?};

\node [pipeBoxNull, below=of stationary]  (null1)  {};
\node [pipeBox, below=of null1]  (hampel)  {Hampel Filter};

\node [pipeNexBox3, below=of hampel] (subcarrierSelection) {Candidate Select (MAD / BoI)};

\node [pipeBox, below=of subcarrierSelection] (lpFilter) {DW Filter};
\node [pipeNexBox3, below=of lpFilter] (peak_detect) {Rate Detect (Peak / PSD)};
\node [pipSig,  below=of peak_detect] (rate) {Rate};
%$\mathbb{C}^{T\times I\items J\times M}$
\path [pipearrow] (csi) --node [text width=2.5cm,midway,right ] {$ \tilde{H}_{i,j,m}[t] \in$ \\ $ \mathbb{C}^{T\times N_T \times N_R \times M}$} (phase_diff);
%\path [pipearrow] (buffer) --node [text width=2.5cm,midway,right ] {$\mathbb{C}^{T\times I\times J\times M}$} (phase_diff);
\path [pipearrow] (phase_diff) --node [text width=2.5cm,midway,right ] {$\angle \tilde{H}_{i,k,m}[t] \in$ \\ $\mathbb{R}^{T\times N_T\times (N_R-1)\times M}$} (stationary);

\path [pipearrow] (phase_diff) -- ++(-0,-0.7)  -- ++(-2, 0) |- (hampel);
\path [pipearrow] (stationary) -- (hampel);
\path [pipearrow] (hampel) --node [text width=2.5cm,midway,right ] {$\angle \tilde{H}_{i,k,m}^{\text{AC}}[t] \in$ \\ $\mathbb{R}^{T\times N_T\times (N_R-1)\times M}$}(subcarrierSelection);

\path [pipearrow] (subcarrierSelection) --node [text width=2.5cm,midway,right ] {$\angle \tilde{H}^*[t] \in \mathbb{R}^{T}$} (lpFilter);

\path [pipearrow] (lpFilter) --node [text width=2.5cm,midway,right ] {$\angle \tilde{H}^*_F[t] \in \mathbb{R}^{T}$} (peak_detect);
\path [pipearrow] (peak_detect) -- (rate);

\end{tikzpicture}

%% file: pics/tikz_ours_long_new2.tex
\tikzstyle{pipeBox} = [rectangle, draw, fill=white!20, node distance=0.7cm, text width=7em, text centered, rounded corners, minimum height=1em, thick]
\tikzstyle{pipeNexBox1} = [rectangle, draw, fill=green!20, node distance=0.7cm, text width=7em, text centered, rounded corners, minimum height=2.5em, thick]
\tikzstyle{pipeNexBox2} = [rectangle, draw, fill=red!20, node distance=0.7cm, text width=7em, text centered, rounded corners, minimum height=1em, thick]
\tikzstyle{pipeNexBox3} = [rectangle, draw, fill=blue!20, node distance=0.7cm, text width=7em, text centered, rounded corners, minimum height=1em, thick]
\tikzstyle{pipSig} = [rectangle, fill=white, node distance=1.0cm, text width=7em, text centered, rounded corners, minimum height=1em, thick]
\tikzstyle{pipearrow} = [draw, -latex',thick]
\tikzstyle{pipe} = [draw=red, -,thick]
\tikzstyle{pipearrow_blue} = [draw=blue, -latex',thick]
\tikzstyle{pipearrow_red} = [draw=red, -latex',thick]

\begin{tikzpicture}[thick,scale=0.6, every node/.style={scale=0.6}]

\node [pipSig] (csi) {CSI};
%\node [pipeBox, below=of csi] (buffer)  {Buffer};
\node [pipeBox, below=of csi]  (phase_diff) {Phase Diff};
\node [pipeBox, below=of phase_diff]  (stationary)  {Stationary?};

\node [pipeNexBox1, below=of stationary] (calibration) {Calibration};
\node [pipeBox, below=of calibration](idft){IDFT};

\node [pipeNexBox3, below=of idft] (pathSelect) {Delay Filter (off / on)};
\node [pipeBox, below=0.1cm of pathSelect](dft){DFT};

\node [pipeBox, below=of dft] (subcarrierSelection) {Candidate Select (BoI)};

\node [pipeBox, below=0.9cm of subcarrierSelection] (lpFilter) {BP Filter};

\node [pipeBox, below=0.9cm of lpFilter] (peak_detect) {Rate Detect (PSD)};

\node [pipSig,  below=0.9cm of peak_detect] (rate) {Rate};

% \path [pipearrow] (csi) --node [text width=2.5cm,midway,right ] {$\mathbb{C}^{T'\times I\times J\times M}$} (buffer);
\path [pipearrow] (csi) --node [text width=2.5cm,midway,right ] {$ \tilde{H}_{i,j,m}[t] \in$ \\ $\mathbb{C}^{T\times N_T \times N_R \times M}$} (phase_diff);
\path [pipearrow] (phase_diff) --node [text width=2.5cm,midway,right ] {$\angle \tilde{H}_{i,k,m}[t] \in$ \\ $\mathbb{R}^{T\times N_T\times (N_R-1)\times M}$} (stationary);

\path [pipearrow] (csi) -- ++(0,-0.7) -- ++(-2,-0) |- (calibration);
\path [pipearrow] (stationary) -- (calibration);
\path [pipearrow] (calibration) --node [text width=2.5cm,midway,right ] {$\hat{H}_{i, j, m}[t] \in $   \\ $\mathbb{C}^{T\times N_T \times N_R \times M}$} (idft);
\path [pipearrow_red] (idft) --node [text width=2.5cm,midway,right ]{$\hat{h}_{i, j, \tau}[t] \in $   \\ $\mathbb{C}^{T\times N_T \times N_R \times M}$} (pathSelect);
\path [pipe] (pathSelect) -- (dft);

\path[pipearrow_blue] (idft) -- ++(0,-0.7) -- ++(-2,-0) --++(0, -3) -- ++(1.6, 0) --++(0, -0.5);
%\path [pipearrow] (calibration) -- ++(0,-0.7) -- ++(-2,-0) -- ++(0,-0.7) --node [text width=2.5cm,midway,right ] {$\hat{h}_{i, j, \tau}[t] \in $   \\ $\mathbb{C}^{T\times N_T \times N_R \times M}$}  ++(0,-4.5) --  (subcarrierSelection);

\path [pipearrow_red] (dft)  --node [text width=2cm,midway,right ] {
$\hat{H'}_{i, j, m}[t] \in $  \\ $\mathbb{C}^{T\times N_T \times N_R \times M}$} (subcarrierSelection);

\path [pipearrow] (subcarrierSelection) --node [text width=2.5cm,midway,right ] {
%$\Phi^*[t] \in \mathbb{R}^{T}$ 
$\color{blue}{h^*[t] \in \mathbb{C}^{T}}$ 
\\ or \\ $\color{red}{H^*[t] \in \mathbb{C}^{T}}$} (lpFilter);

\path [pipearrow] (lpFilter) --node [text width=2.5cm,midway,right] {
%$\Phi^*_F[t] \in \mathbb{R}^{T}$
$\color{blue}{h^*_F[t] \in \mathbb{C}^{T}}$
\\ or \\ $\color{red}{H^*_F[t] \in \mathbb{C}^{T}}$} (peak_detect);

\path [pipearrow] (peak_detect) -- (rate);

\end{tikzpicture}

%% file: sec/experiment_new.tex
\section{Experimental Study}
\label{sec:experiment}
In this section, we evaluate the accuracy of the two systems and their variants that are described in Section \ref{sec:System} on the same data set.

\subsection{Data Set}
We use the data set provided with \cite{hillyard_experience_2018}. 
This data set records the \gls{csi} from a WiFi MIMO OFDM system that is running overnight, while patients are sleeping between the transmitter and the receiver.
The WiFi system consists of two TX antennas and two RX antennas and operates in the $5$ GHz ISM band with $114$ subcarriers in a $40$~MHz bandwidth. 
To make the dataset structure as similar as possible to the dataset used in~\cite{wang_phasebeat_2017}, where there are one TX and two RX antennas operating in a bandwidth of $20$~MHz with $30$~subcarriers, we only keep the measured data on one TX antenna and the upper half frequency band. 
We pick every second subcarrier, which yields~$29$ subcarriers.
\gls{csi} data is sampled for each antenna pair at a sampling rate of $9.9$ Hz. 
By sliding a window with a size of $30$ seconds, we obtain the \gls{csi} $\tilde{H}_{i,j,m}[t] \in \mathbb{C}^{297 \times 1\times 2 \times 29}$.
For each new estimation iteration, we slide the time window by~$1$ second.
To check the estimation performance for different ground truth breathing frequencies, we picked the patients with indices~$\{00, 03, 06, 07, 10, 11, 12, 16\}$, for which PhaseBeat shows a good absolute estimation error below 1.2 beats per minute (bpm) for more than $60\%$ of the considered snapshots.
This set of patients' breathing rate covers the range from~$0.2$ to~$0.5$~Hz, i.e., $12$ to $30$~bpm.

The data set also offers PSG data as a reference, from which we obtain the ground truth breathing rate  with the method described in \cite{hillyard_experience_2018}.

%The data is first pre-processed as described in \cite{xie_precise_2019}, in order to fix the phase bonding problem, such that the phase shift caused by combining the bands in the measurement 

\subsection{Performance Analysis and Comparison}

Since the PhaseBeat system in \cite{wang_phasebeat_2017} is only tested for breathing rate range from $0.2$ to $0.3$~Hz, we first compare the estimation error for the case where the ground truth breathing rate is within the range from $0.2$ to $0.3$~Hz.
%\begin{figure}[t]
%\centering
%\input{./pics/tikz_plot_result_est_short.tex}
%\caption{Estimated breathing rate compared to reference breathing rate, patient index $03$, whose breathing rate is in the range from $12.4$ to $12.9$ %bpm, within $1$ minutes.}
%\label{fig:result_est}
%\end{figure}
%\begin{figure}[t]
%\centering
%\input{./pics/tikz_plot_result_est2_short.tex}
%\caption{Estimated breathing rate compared to reference breathing rate, patient index $07$, whose breathing rate is in the range from $18.8$ to $19.5$ %bpm, within $1$ minutes.}
%\label{fig:result_est2}
%\end{figure}
%\begin{figure}[t]\centering
%\subcaptionbox{Patient index $03$, breathing rate is in the range from $13.1$ to $14.3$ bpm. \label{fig:result_est}}{
%\input{./pics/tikz_plot_result_est_short.tex}
%}\quad
%\subcaptionbox{Patient index $07$, breathing rate is in the range from $18.7$ to $19.9$ bpm. \label{fig:result_est2}}{
%\input{./pics/tikz_plot_result_est2_short.tex}
%}
%\caption{Estimated breathing rate compared to reference breathing rate.}
%\end{figure}
%\begin{figure}[t]\centering
%\subcaptionbox{Displacement sensitivity \label{fig:subsel1}}{
%\input{./pics/tikz_subsel_1.tex}
%}\quad
%\subcaptionbox{Stability against noise \label{fig:subsel2}}{
%\input{./pics/tikz_subsel_2.tex}
%}
%\caption{CSI on the complex plane for different subcarriers.}
%\end{figure}
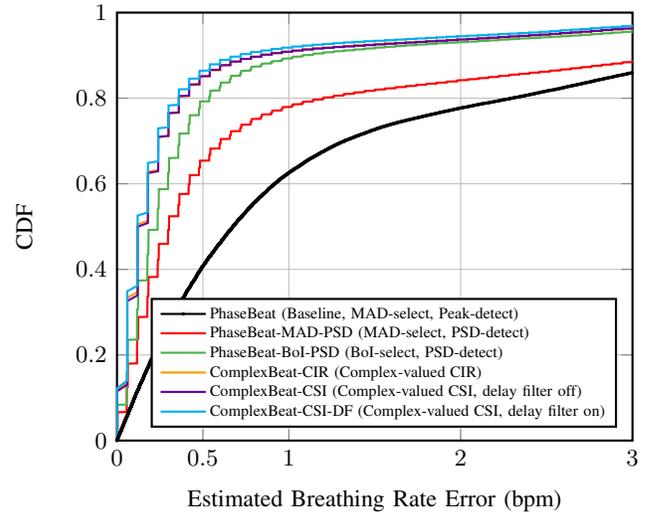
\begin{figure}[t]
\centering
\input{./pics/tikz_plot_result_cdf.tex}
\caption{CDF of estimation error with breathing rate range from $0.2$ to $0.3$~Hz, where the CDF of ComplexBeat-CIR (\ref{leg:ComplexBeat-CIR}) is almost covered by the CDF of ComplexBeat-CSI  (\ref{leg:ComplexBeat-CSI}). }
\label{fig:cdf_small}
\end{figure}
The CDFs of the absolute estimation errors are shown in Fig.~\ref{fig:cdf_small}. 
Two variants of PhaseBeat, PhaseBeat-MAD-PSD and PhaseBeat-BoI-PSD, show better performance than the original PhaseBeat. 
The combination of the \gls{boi}-based selection method and the \gls{psd}-rate detection method shows the best performance, where $79\%$ of the estimation error is less than $0.5$~bpm. 
Our ComplexBeat system outperforms PhaseBeat and its variants.
%Although ComplexBeat-P has worse performance compared to PhaseBeat-BS, with our calibration method, it provides the possibility to be potentially further processed in the delay domain. 
ComplexBeat-CIR and ComplexBeat-CSI achieve similar performance.
With filtering in the delay domain, and the candidate selected in the frequency domain, ComplexBeat-CSI-DF shows the best result, where $86\%$ of the estimation errors are below $0.5$~bpm. 
%As shown in the CDF curve for ComplexBeat-PD, with the delay filter activated, ComplexBeat-PD reaches slightly better performance than PhaseBeat-BF,  where $76\%$ of the estimation error is within the range of $0.5$~bpm. 
%The ComplexBeat-CD shows the best performance among all methods, where $82\%$ of the estimation error is within the range of $0.5$~bpm. 
%By checking the periodicity of the filtered and complex-valued \gls{csi}, we achieve the highest accuracy compared to all the other tested methods.
%\textcolor{red}{Compared to PhaseBeat, ComplexBeat-CSI-DF is also more efficient by $94\%$ in running time consumption. The main reason is that ComplexBeat and it variants avoid using the Hample filter.}
We also calculated the CDF of the estimation error when the ground truth breathing rate is between $0.2$ and $0.5$~Hz.
In this case, ComplexBeat-CSI-DF still shows the best performance among all the systems.
Due to space limitations, the CDF plot for this case is not included in this paper.
%Thus it is essential to extract the CSI phase component for respiration estimation

%% file: pics/tikz_plot_result_cdf.tex
\begin{tikzpicture}
%%%%%%%%%%IMPORTANT%%%%%%%%%%%%%%%%%%%%%%%%%
% V1 WITHOUT point iteration 3,5
\begin{axis}[thick,
%scale=0.75, 
%every node/.style={scale=0.75}, 
%width=8cm,height=7cm, 
legend style={nodes={scale=0.6, transform shape}}, 
legend cell align=left, xlabel=Estimated Breathing Rate Error (bpm), ylabel=CDF, grid=both,  
label style={font=\small},
tick label style={font=\small}, 
xtick={0,0.5,1,2,3}, 
ymin=0, ymax=1, 
xmin=0, xmax=3, 
legend pos=south east, 
%ymin=1e-7
]
\addplot [thick, color=black,
    mark=*,  
    mark size = 0.2pt]
    table [x=qe, y=pe, col sep=comma] {./csv/cdf_plot_diff0.csv};\addlegendentry{PhaseBeat (Baseline, MAD-select, Peak-detect)}
\addplot [thick, color=red,
    mark=none,  
    mark size = 0.2pt]
    table [x=qe, y=pe, col sep=comma] {./csv/cdf_plot_diff.csv};\addlegendentry{PhaseBeat-MAD-PSD (MAD-select, PSD-detect)}
\addplot [thick, color=green!40!gray,
    mark=none, 
    mark size = 0.2pt]
    table [x=qe, y=pe, col sep=comma] {./csv/cdf_plot_diff2.csv};\addlegendentry{PhaseBeat-BoI-PSD (BoI-select, PSD-detect)}
%\addplot [thick, color=yellow!40!gray,
%    mark=none, 
%    mark size = 0.2pt]
%    table [x=qe, y=pe, col sep=comma] {./csv/cdf_plot_cp0.csv};\addlegendentry{ComplexBeat-P (Phase-only, delay filter off)}
\addplot [thick, color=red!40!yellow,
    mark=none, 
    mark size = 0.2pt]
    table [x=qe, y=pe, col sep=comma] {./csv/cdf_plot_cir.csv};\addlegendentry{ComplexBeat-CIR (Complex-valued CIR)}\label{leg:ComplexBeat-CIR}
%\addplot [thick, color=red!40!blue,
%    mark=none, 
%    mark size = 0.2pt]
%    table [x=qe, y=pe, col sep=comma] {./csv/cdf_plot_cp.csv};\addlegendentry{ComplexBeat-PD (Phase-only, delay filter on)}
\addplot [thick, color=red!40!blue,
    mark=none, 
    mark size = 0.2pt]
    table [x=qe, y=pe, col sep=comma] {./csv/cdf_plot_cn.csv};\addlegendentry{ComplexBeat-CSI (Complex-valued CSI, delay filter off)}\label{leg:ComplexBeat-CSI}
\addplot [thick, color=cyan,
    mark=none, 
    mark size = 0.2pt]
    table [x=qe, y=pe, col sep=comma] {./csv/cdf_plot_c.csv};\addlegendentry{ComplexBeat-CSI-DF (Complex-valued CSI, delay filter on)}
%\addplot [thick, color=blue,
%    mark=oplus*, dashed,
%    mark size = 0.2pt]
%    table [x=qe, y=pe, col sep=comma] {./csv/cdf_plot_path_filt.csv};\addlegendentry{Estimation System $3$ (filtered)}
\end{axis}
\end{tikzpicture}

%% file: sec/conclusions.tex
\section{Conclusions}
%In this paper, we illustrated that it is beneficial to use CSI phase components that are more related and sensitive to the human chest displacement for breathing rate estimation. 
In this paper, we show that robust breathing rate estimation from \gls{csi} relies on careful selection of the best \gls{csi} component, especially in the delay domain.
%We introduced the Normalized-TV-MAD-based candidates selection criterion to select CSI phase components and compared it to the MAD-based candidate selection method.
%Experiments show that the Normalized-TV-MAD-based method performs better than MAD-based methods in terms of breathing rate estimation accuracy.
To this end, we introduce a dominant path-based \gls{csi} amplitude calibration method and a linear component-based \gls{csi} phase calibration method for pre-processing of the \gls{csi} data.
The calibration methods combined with the delay domain filtering leads to a better \gls{csi} extraction compared to PhaseBeat and its variants.
Furthermore, we show that the breathing rate can be estimated directly from calibrated complex-valued \gls{csi} or \gls{cir}.
Compared to observing only the phase values, the estimation from complex-valued \gls{csi} achieves the highest accuracy.
%Using this calibration method together with the delay filter method we proposed, the breathing rate estimation achieves a high estimation accuracy.

%In future work, we should implement those systems for breathing rate estimation in channels with increased disturbance and noise for multiple paths, and observe the performance of different CSI processing and extraction methods, especially, we will study the robustness of the delay domain filter against the increasing number of paths, that contains the breathing information.